\documentclass[journal,12pt,draftclsnofoot,onecolumn,twoside]{IEEEtran}
\usepackage{ifpdf}
 \ifpdf
     \usepackage[pdftex]{graphicx}
 \else
     \usepackage[dvips]{graphicx}
 \fi
\graphicspath{{figures/}}

\usepackage{cite}
\usepackage{epstopdf}					% convert eps figure automatically to pdf for pdflatex
\usepackage[cmex10]{amsmath}
\usepackage{amssymb,amsfonts}
\usepackage{algorithm}
\usepackage{algpseudocode}
\usepackage{mdwmath}
\usepackage{mdwtab}

\usepackage[belowskip=-14pt,aboveskip=-4pt,small]{caption}
\begin{document}
%\title{Joint Bit and Power Loading for Multicarrier Systems with Total Power Constraint}
\title{Constrained Joint Bit and Power Allocation for Multicarrier Systems}

%\author{Author names\\
%\IEEEauthorblockA{Faculty of Engineering and Applied Science, Memorial University,
%St. John's, NL, Canada\\
%\IEEEauthorrefmark{2} Communications Research Centre, Ottawa, ON, Canada\\
%Email: \{\}@mun.ca, kareem.baddour@crc.ca}
%}
\author{{Ebrahim Bedeer, %\IEEEauthorrefmark{1},
Octavia A. Dobre, %\IEEEauthorrefmark{1},
Mohamed H. Ahmed, and
Kareem E. Baddour \IEEEauthorrefmark{2}}\\
\IEEEauthorblockA{Faculty of Engineering and Applied Science, Memorial University of Newfoundland,
St. John's, NL, Canada\\
\IEEEauthorrefmark{2} Communications Research Centre, Ottawa, ON, Canada\\
Email: \{e.bedeer, odobre, mhahmed\}@mun.ca, kareem.baddour@crc.ca}
}
\maketitle

\begin{abstract}
%\boldmath
This paper proposes a novel low complexity joint bit and power suboptimal allocation algorithm for multicarrier systems operating in fading environments. The algorithm jointly maximizes the throughput and minimizes the transmitted power, while guaranteeing a target bit error rate (BER) per subcarrier and meeting a constraint on the total transmit power. Simulation results are described that illustrate the performance of the proposed scheme and demonstrate its superiority when compared to the algorithm in [4] with similar or reduced computational complexity. Furthermore, the results show that the performance of the proposed suboptimal algorithm approaches that of an optimal exhaustive search with significantly lower computational complexity.
%Cognitive Radio (CR) allows cognitive users (CUs) to either opportunistically access voids in existing users (EUs) frequency bands/time slots; or share the spectrum with EUs as long as no internecine interference incurs to both users. In this paper, we consider the latter case, where a CU coexists with a narrowband (NB) EU in a frequency selective fading environment, and present an adaptive resource allocation algorithm to maximizes the CU throughput while guaranteeing the CU bit error rate (BER) below a predefined threshold. Moreover, we study the effect of imperfect channel estimation on the presented algorithm.
%
%This paper presents an optimal bit and power loading low complexity algorithm for multicarrier communication systems operating in frequency selective fading environment. This algorithm, jointly, minimizes the transmitted power, as well as, maximizes the throughput while guaranteeing a target bit error rate (BER) per subcarrier. Closed-form expressions for optimal bit and power distributions per subcarrier are derived. Simulation results indicate the superiority of the proposed algorithm over various bit and power loading algorithms known in the literature.
\end{abstract}


% \fi
%\IEEEpeerreviewmaketitle

\section{Introduction}

Multicarrier modulation is recognized as a robust and efficient transmission technique, as evidenced by its consideration for diverse communication systems and adoption by several wireless standards \cite{fazel2008multi, mahmoud2009ofdm}. The performance of multicarrier communication systems can be significantly improved by dynamically adapting the transmission parameters, such as power, constellation size, symbol rate, coding rate/scheme, or any combination of these, according to the channel conditions or the wireless standard specifications \cite{krongold2000computationally, wyglinski2005bit, fox1966discrete, papandreou2005new, liu2009adaptive, bedeer2012adaptiveRWS, song2002joint, bedeer2012jointVTC}.
%\cite{hughes1988ensemble, de1998optimal, levin2001complete, wyglinski2005bit, fox1966discrete, song2002joint, sonalkar2000efficient, papandreou2005new, liu2009adaptive, goldfeld2002minimum, krongold2000computationally}.

Generally speaking, the problem of optimally loading bits and power per subcarrier can be categorized into two main classes: \textit{rate maximization} (RM) and \textit{margin maximization} (MM) \cite{krongold2000computationally, wyglinski2005bit, fox1966discrete, papandreou2005new, liu2009adaptive}. For the former, the objective is to maximize the achievable data rate \cite{krongold2000computationally, wyglinski2005bit}, while for the latter the objective is to maximize the achievable system margin \cite{papandreou2005new, liu2009adaptive} (i.e., minimizing the total transmit power given a target data rate). Most of the prior work has focused on optimizing either the RM or the MM problem separately. Krongold \textit{et al.} \cite{krongold2000computationally} presented a computationally efficient algorithm to maximize the throughput using a look-up table search and the Lagrange multiplier bisection method. The algorithm converges faster to the optimal solution when compared to other allocation schemes. In \cite{wyglinski2005bit}, Wyglinski \textit{et al.} proposed an incremental bit loading algorithm to maximize the throughput while guaranteeing a target mean BER. The algorithm nearly achieves the optimal solution given in \cite{fox1966discrete} but with lower complexity. On the other hand,  Papandreou and Antonakopoulos \cite{papandreou2005new} presented an efficient bit loading algorithm to minimize the transmit power that converges faster to the same bit allocation as the discrete optimal bit-filling and bit-removal methods. The algorithm exploits the differences between the subchannel gain-to-noise ratios in order to determine an initial bit allocation and then performs a multiple bit insertion or removal loading procedures to achieve the requested target rate. In \cite{liu2009adaptive}, Liu and Tang proposed a low complexity power loading algorithm that aims to minimize the transmit power while guaranteeing a target average BER.
Song \textit{et~al.} \cite{song2002joint} proposed an iterative joint bit loading and power allocation algorithm based on \textit{statistical} channel conditions to meet a target BER, i.e., the algorithm loads bits and power per subcarrier based on long-term frequency domain channel conditions, rather than \textit{instantaneous} channel conditions, as in \cite{krongold2000computationally, wyglinski2005bit, fox1966discrete, papandreou2005new, liu2009adaptive}. The algorithm marginally improves performance when compared to conventional multicarrier systems. The authors conclude that their algorithm is not meant to compete with its counterparts that adapt according the instantaneous channel conditions.
In \cite{bedeer2012jointVTC}, the authors proposed a novel algorithm that jointly maximizes the throughput and minimizes the transmit power, while guaranteeing a target average BER.

In emerging wireless communication systems, different requirements are needed. For example, minimizing the transmit power is prioritized when operating in interference-prone shared spectrum environments or in proximity to other frequency-adjacent users. On the other hand, maximizing the throughput is favoured if sufficient guard bands exist to separate users. This motivates us to jointly optimize the RM and MM problems, by introducing a weighting factor that reflects the importance of the competing throughput and power objectives.

%\vspace{-8pt}
In this paper, we propose a suboptimal algorithm that jointly maximizes the throughput and minimizes the total transmit power, subject to constraints on the BER per subcarrier and the total transmit power. Limiting the total transmit power reduces the interference to existing users, which is crucial in various wireless networks, including cognitive radio environments. Moreover, including the total subcarrier power in the objective function is especially desirable as it minimizes the transmit power even when the power constraint is ineffective, which occurs at smaller signal-to-noise ratios (SNR). Simulation results show that the proposed algorithm outperforms Wyglinski's algorithm \cite{wyglinski2005bit} with similar or reduced computational complexity. The results also  indicate that the proposed suboptimal algorithm's performance approaches that of an exhaustive search for the optimal discrete allocations, with significantly reduced computational complexity.

%\vspace{-2pt}
The remainder of the paper is organized as follows. Section \ref{sec:opt} introduces the proposed joint bit and power allocation algorithm. Simulation results are presented in Section \ref{sec:sim}, while conclusions are drawn in Section \ref{sec:conc}.

\section{Proposed Algorithm} \label{sec:opt}
\subsection{Optimization Problem Formulation}
A multicarrier communication system decomposes the signal bandwidth into a set of $N$ orthogonal narrowband subcarriers of equal bandwidth. Each subcarrier $i$ transmits $b_i$ bits using power $\mathcal{P}_i$, $i = 1, ..., N$. A delay- and error-free feedback channel is assumed to exist between the transmitter and receiver for reporting channel state information.

In order to minimize the total transmit power and maximize the throughput subject to BER and total power constraints, the optimization problem is  formulated as
\setlength{\arraycolsep}{0.0em}
\begin{equation}
\underset{\mathcal{P}_i}{\textup{Minimize}} \; \sum_{i = 1}^{N}\mathcal{P}_i \quad \textup{and} \quad  \underset{b_i}{\textup{Maximize}} \; \sum_{i = 1}^{N}b_i, \nonumber
\end{equation}
\begin{eqnarray}
\textup{subject to}  & \quad & \textup{BER}_i \leq \textup{BER}_{th,i}, \nonumber \\
 										 & \quad & \sum_{i = 1}^{N}\mathcal{P}_i \leq \mathcal{P}_{th}, \: \: \: i = 1, ..., N,  \label{eq:ob1}
\end{eqnarray}
where $\textup{BER}_i$ and $\textup{BER}_{th,i}$ are the BER and threshold value of BER per subcarrier $i$, respectively, and $\mathcal{P}_{th}$ is the total transmit power threshold. An approximate expression for the BER per subcarrier $i$ in case of $M$-ary QAM is given by\footnote[1]{This expression is tight within 1 dB for BER $\leq$ $10^{-3}$.} \cite{liu2009adaptive}
\setlength{\arraycolsep}{0.0em}
\begin{eqnarray}
\textup{BER}_i &{} \approx  {}& 0.2 \: \textup{exp}\left ( -1.6 \frac{\mathcal{P}_i}{(2^{b_i} - 1)} \frac{\left | \mathcal{H}_i \right |^2}{\sigma^2_n } \right ), \label{eq:BER}
%\textup{BER}_i &{} \approx  {}& 0.2 \: \textup{exp}\left ( \frac{- \mathcal{C}_i \mathcal{P}_i}{2^{b_i} - 1} \right ), \label{eq:BER}
\end{eqnarray}
where $ \mathcal{H}_i $ is the channel gain of subcarrier $i$ and $\sigma^2_n$ is the variance of the additive white Gaussian noise (AWGN).

The multi-objective optimization function in (\ref{eq:ob1}) can be rewritten as a linear combination of multiple objective function as follows
\setlength{\arraycolsep}{0.0em}
\begin{eqnarray}
\underset{\mathcal{P}_i, b_i}{\textup{Minimize}} &{} \:\: {}& \mathcal{F}(\textbf{\textit{P}},  \textbf{\textit{b}}) = \: \left\{\alpha \sum_{i = 1}^{N}\mathcal{P}_i - (1-\alpha)\sum_{i = 1}^{N}b_i\right\}, \nonumber \\
\textup{subject to} &{} \:\: {}& g_j(\mathcal{P}_i, b_i) \leq 0, i = 1, ..., N, j = 1, ..., N+1, \label{eq:ineq_const}
\end{eqnarray}
where $\alpha$ ($0 < \alpha < 1$) is a constant whose value indicates the relative importance of one objective function relative to the other, $\textbf{\textit{P}} = [\mathcal{P}_1, ...,  \mathcal{P}_N]^T$ and $\textbf{\textit{b}} = [b_1, ..., b_N]^T$ are the \textit{N}-dimensional power and bit distribution vectors, respectively, with $[.]^T$ denoting the transpose operation, and $g_j(\mathcal{P}_i, b_i)$\footnote[2]{Note that $g_j(\mathcal{P}_i, b_i)$ is a function of $\mathcal{P}_i$ and $b_i$ for $i = j$. When $j = N+1$, it is a function of $\mathcal{P}_i$, $i$ = 1, ..., $N$.} is the set of $N+1$ constraints given by
\setlength{\arraycolsep}{0.0em}
\begin{equation}
g_j(\mathcal{P}_i, b_i) = \left\{\begin{matrix}
\hspace{-1cm} 0.2 \: \textup{exp}\left ( \frac{- 1.6 \: \mathcal{C}_i \mathcal{P}_i}{2^{b_i} - 1} \right ) - \textup{BER}_{th,i} \leq 0, \\ \hspace{4.5cm} j = 1, ..., N  \\
 \sum_{i = 1}^{N}\mathcal{P}_i - \mathcal{P}_{th} \leq 0, \hfill j = N+1
\end{matrix}\right.
\label{eq:cons}
\end{equation}
where $\mathcal{C}_i = \: \frac{\left | \mathcal{H}_i \right |^2}{\sigma^2_n}$ is the channel-to-noise ratio for subcarrier~$i$.

\subsection{Optimization Problem Analysis and Solution}
The problem in (\ref{eq:ineq_const}) can be solved by applying the method of Lagrange multipliers. Accordingly,  the inequality constraints in (\ref{eq:cons}) are transformed to equality constraints by adding non-negative slack variables, $\mathcal{Y}_j^2$, $j$ = 1, ..., $N+1$ \cite{griva2009linear}. Hence, the constraints are rewritten as
\setlength{\arraycolsep}{0.0em}
\begin{eqnarray}
%\textup{Minimize} &{} \quad {}& f(\textbf{\textit{P}},\textbf{\textit{b}}) = \: \left\{\alpha \sum_{i = 1}^{N}\mathcal{P}_i - (1-\alpha)\sum_{i = 1}^{N}b_i\right\}\\
\mathcal{G}_j(\mathcal{P}_i, b_i, \mathcal{Y}_j) &{} = {}& g_j(\mathcal{P}_i, b_i) + \mathcal{Y}_j^2 = 0, \: j = 1, ..., N+1,
\label{eq:eq_const}
\end{eqnarray}
and further, the Lagrange function $\mathcal{L}$ is expressed as
\setlength{\arraycolsep}{0.0em}
\begin{eqnarray}
\mathcal{\mathcal{L}}(\textbf{\textit{P}}, \textbf{\textit{b}}, \textbf{\textit{Y}},   \mathbf{\Lambda}) &{} = {}& \mathcal{F}(\textbf{\textit{P}}, \textbf{\textit{b}}) + \sum_{j = 1}^{N+1} \lambda_j \: \mathcal{G}(\mathcal{P}_i, b_i, \mathcal{Y}_j), \nonumber \\
 &{} = {}& \alpha \sum_{i = 1}^{N}\mathcal{P}_i - (1-\alpha)\sum_{i = 1}^{N}b_i \nonumber \\
& & + \sum_{i = 1}^{N} \lambda_i\:\Bigg[ 0.2 \: \textup{exp}\left ( \frac{- 1.6 \: \mathcal{C}_i \mathcal{P}_i}{2^{b_i} - 1} \right ) - \textup{BER}_{th,i} \nonumber \\
& & \hspace{3.9cm}+ \mathcal{Y}_i^2\Bigg] \nonumber \\
& & + \lambda_{N+1}\: \Bigg[ \sum_{i = 1}^{N}\mathcal{P}_i - \mathcal{P}_{th} + \mathcal{Y}_{N+1}^2  \Bigg], \label{eq:L}
\end{eqnarray}
where $\mathbf{\Lambda} = [\lambda_1, ..., \lambda_{N+1}]^T$ and $\textbf{\textit{Y}} = [\mathcal{Y}_1^2, ..., \mathcal{Y}_{N+1}^2]^T$ are the vectors of Lagrange multipliers and slack variables, respectively. A stationary point can be found when $\nabla \mathcal{L}(\textbf{\textit{P}}, \textbf{\textit{b}}, \textbf{\textit{Y}}, \mathbf{\Lambda}) = 0$ (where $\nabla$ denotes the gradient), which yields
\setlength{\arraycolsep}{0.0em}
\begin{eqnarray}
\frac{\partial \mathcal{L}}{\partial \mathcal{P}_i} &{}  =  {}& \: \alpha - 0.2 \: \lambda_i \frac{1.6 \: \mathcal{C}_i}{2^{b_i}-1} \: \textup{exp}\left ( \frac{- 1.6 \: \mathcal{C}_i \mathcal{P}_i}{2^{b_i} - 1} \right ) \nonumber \\ & & \hspace{4.5cm} + \lambda_{N+1} = 0,\label{eq:eq1}\\
\frac{\partial \mathcal{L}}{\partial b_i} &{}  = {}& \: -(1 - \alpha) + 0.2 \ln (2) \: \lambda_i \frac{1.6 \: \mathcal{C}_i \mathcal{P}_i 2^{b_i}}{(2^{b_i}-1)^2} \: \nonumber \\ & & \hspace{2.6cm} \times \: \textup{exp}\left ( \frac{- 1.6 \: \mathcal{C}_i \mathcal{P}_i}{2^{b_i} - 1} \right )  = 0,\label{eq:eq2}\\
\frac{\partial \mathcal{L}}{\partial \lambda_i} &{}  = {}& \hspace{0.1cm}  0.2 \: \textup{exp}\left ( \frac{- 1.6 \: \mathcal{C}_i \mathcal{P}_i}{2^{b_i} - 1} \right ) - \textup{BER}_{th,i} + \mathcal{Y}_i^2= 0, \label{eq:eq3}\\
\frac{\partial \mathcal{L}}{\partial \lambda_{N+1}} &{}  = {}& \quad \sum_{i = 1}^{N}\mathcal{P}_i - \mathcal{P}_{th} + \mathcal{Y}_{N+1}^2 = 0, \label{eq:eq3_1}\\
\frac{\partial \mathcal{L}}{\partial \mathcal{Y}_i} &{}  = {}& \quad 2\lambda_i \mathcal{Y}_i = 0, \label{eq:eq4}\\
\frac{\partial \mathcal{L}}{\partial \mathcal{Y}_{N+1}} &{}   = {}& \quad 2\lambda_{N+1} \mathcal{Y}_{N+1} = 0. \label{eq:eq4_1}
\end{eqnarray}
It can be seen that (\ref{eq:eq1}) to (\ref{eq:eq4_1}) represent $4N+2$ equations in the $4N+2$ unknown components of the vectors $\textbf{\textit{P}}, \textbf{\textit{b}}, \textbf{\textit{Y}}$,  and $\mathbf{\Lambda}$. Equation (\ref{eq:eq4}) implies that either $\lambda_i$ = 0 or $\mathcal{Y}_i$ = 0, $i$ = 1, ..., $N$, while (\ref{eq:eq4_1}) implies that either $\lambda_{N+1}$ = 0 or $\mathcal{Y}_{N+1}$ = 0. Accordingly, four possible solutions exist, as follows:

--- \textit{Solutions \MakeUppercase{\romannumeral 1} \& \MakeUppercase{\romannumeral 2}}:
Choosing $\lambda_i$ = 0, $i$ = 1, ..., $N$, and $\mathcal{Y}_{N+1}$ or $\lambda_{N+1}$ = 0, results in an underdetermined system of $N+1$ equations in $3N+1$ unknowns; hence, no unique solution can be reached.

--- \textit{Solution \MakeUppercase{\romannumeral 3}}: Choosing $\mathcal{Y}_i$ = 0, $i$ = 1, ..., $N$, and $\mathcal{Y}_{N+1}$ = 0, results in $3N+1$ nonlinear equations in $3N+1$ unknowns that represent the optimal solution if the total transmit power constraint is active.
%However, setting $\mathcal{Y}_{N+1}$ = 0 in (\ref{eq:eq3_1}) forces the sum of optimal power allocation $\mathcal{P}_i$ to always equal the power constraint, which is not feasible. Furthermore, it results in a negative $\lambda_{N+1}$, which violates the Karush-Khun-Tucker (KKT) optimality conditions, as discussed later.

--- \textit{Solution \MakeUppercase{\romannumeral 4}}:
Choosing $\mathcal{Y}_i$ = 0, $i$ = 1, ..., $N$, and $\lambda_{N+1}$ = 0, results in $3N+1$ nonlinear equations in $3N+1$ unknowns that represent the optimal solution if the total transmit power constraint is inactive.
%cannot be solved analytically\footnote[3]{A possible numerical solution is not discussed here due to space limitation; this will be presented in future work. Similar results were obtained with the numerical solution.}.

We resort to a low complex suboptimal solution, which is obtained as follows. The constraint on the total transmit power in (\ref{eq:eq3_1}) is first relaxed, and the optimal solution of (\ref{eq:eq1}) to (\ref{eq:eq3}) is found. This provides the initial values for $\textbf{\textit{b}}, \textbf{\textit{P}}$, denoted by $\textbf{\textit{b}}^*, \textbf{\textit{P}}^*$, to be used with the suboptimal algorithm. Then, the final bit and power distributions are reached in an iterative manner to meet the power and BER constraints. The suboptimal algorithm will be presented in Section \MakeUppercase{\romannumeral 2}-\textit{C}; the optimal solution for the initial bit and power distributions while relaxing the power constraint is provided below.

--- \textit{Calculation of the initial optimal bit and power distributions, $\textbf{\textit{b}}^*, \textbf{\textit{P}}^*$}: In solution \textit{\MakeUppercase{\romannumeral 4}}, by relaxing the power constraint in (\ref{eq:eq3_1}), we obtain $3N$ equations in the $3N$ unknowns $\textbf{\textit{P}}, \textbf{\textit{b}}$, and $\mathbf{\Lambda}$ $(\lambda_{N+1} = 0)$ that can be solved analytically, as follows.

From (\ref{eq:eq1}) and (\ref{eq:eq2}), we can relate $\mathcal{P}_i$ and $b_i$ as
\setlength{\arraycolsep}{0.0em}
\begin{eqnarray}
\mathcal{P}_i &{} = {}& \frac{1- \alpha}{\alpha \ln(2)}(1 - 2^{-b_i}), \label{eq:eq8}
\end{eqnarray}
with $\mathcal{P}_i \geq 0$ if and only if $b_i \geq 0$. By substituting (\ref{eq:eq8}) into (\ref{eq:eq3}), one obtains the solution
\setlength{\arraycolsep}{0.0em}
\begin{eqnarray}
b_i^* = \frac{1}{\log(2)}\log\Bigg(- \frac{1-\alpha }{\alpha \ln(2)} \frac{1.6 \: \mathcal{C}_i}{\ln(5 \: \textup{BER}_{th,i})}\Bigg). \label{eq:eq10}
\end{eqnarray}
Consequently, from (\ref{eq:eq8}) one gets
\setlength{\arraycolsep}{0.0em}
\begin{eqnarray}
\mathcal{P}_i^* = \frac{1-\alpha }{\alpha \ln(2)}\Bigg( 1 - \Big(- \frac{1-\alpha }{\alpha \ln(2)} \frac{1.6 \: \mathcal{C}_i}{\ln(5 \: \textup{BER}_{th,i})} \Big)^{-1} \Bigg). \label{eq:eq12}
\end{eqnarray}
Since (\ref{eq:BER}) is valid for $M$-ary QAM, $b_i$ should be greater than 2. From (\ref{eq:eq10}), to have $b_i \geq 2$, the channel-to-noise ratio per subcarrier, $\mathcal{C}_i$, must satisfy the condition
\setlength{\arraycolsep}{0.0em}
\begin{eqnarray}
\mathcal{C}_i \geq - \frac{4}{1.6} \: \frac{\alpha \ln(2)}{1-\alpha} \: \ln(5\:\textup{BER}_{th,i}), \qquad i = 1, ..., N. \label{eq:condition}
\end{eqnarray}

The obtained solution represents a minimum of $\mathcal{F}(\textbf{\textit{P}},\textbf{\textit{b}})$ if the KKT conditions are satisfied \cite{griva2009linear}. Given that our stationary point ($b_i^*$, $\mathcal{P}_i^*$) in (\ref{eq:eq10}) and (\ref{eq:eq12}) exists at $\mathcal{Y}_i = 0$, $i$ = 1, ..., $N$, the KKT conditions can be written as
\setlength{\arraycolsep}{0.0em}
\begin{eqnarray}
\frac{\partial \mathcal{F}}{\partial \mathcal{P}_i} + \sum_{\rho = 1}^{N}\lambda_{\rho} \: \frac{\partial g_{\rho}}{\partial \mathcal{P}_i} &{} = {}& 0, \label{eq:KH1}\\
\frac{\partial \mathcal{F}}{\partial b_i} +  \sum_{\rho = 1}^{N}\lambda_{\rho} \: \frac{\partial g_{\rho}}{\partial b_i} &{} = {}& 0, \label{eq:KH2}\\
\lambda_{\rho} &{} > {}& 0, \hspace{1cm} \rho = 1, ..., N. \label{eq:KH3}
\end{eqnarray}
One can easily prove that these conditions are fulfilled, as follows.

--- \textit{Proof of} (\ref{eq:KH1})-(\ref{eq:KH3}): From (\ref{eq:eq1}), one finds
\setlength{\arraycolsep}{0.0em}
\begin{eqnarray}
\lambda_i = \alpha \Bigg[ 0.2 \: \frac{1.6 \: C_i}{2^{b_i}-1} \: \textup{exp}\Big( \frac{-1.6 \: C_i P_i}{2^{b_i}-1}  \Big)  \Bigg]^{-1}, \label{eq:lambda}
\end{eqnarray}
which is positive for all values of $i$, and hence it satisfies (\ref{eq:KH3}). Moreover, by substituting (\ref{eq:eq10}), (\ref{eq:eq12}), and (\ref{eq:lambda}) in (\ref{eq:KH1}) and (\ref{eq:KH2}), one can easily verify that the KKT conditions are satisfied. Note that $\textbf{\textit{b}}^*, \textbf{\textit{P}}^*$ represent an optimal solution when there is no constraint on the total transmit power. \hfill$\blacksquare$

\subsection{Proposed Joint Bit and Power Suboptimal Allocation Algorithm}
The solution ($\textbf{\textit{b}}^*, \textbf{\textit{P}}^*$) given in (\ref{eq:eq10}) and (\ref{eq:eq12}) is obtained for $\lambda_{N+1}$ = 0, which basically means that no constraint on the total transmit power is considered for the problem formulated in (\ref{eq:ineq_const}). To consider such a constraint, we propose a suboptimal algorithm whose idea is as follows. The total power $\sum_{i = 1}^{N}\mathcal{P}_i$ is calculated based on (\ref{eq:eq12}) and checked against the threshold value $\mathcal{P}_{th}$. If less than the threshold, then the power constraint is met; otherwise, the power $\Delta \mathcal{P}_i = \mathcal{P}_i(b_i) - \mathcal{P}_i(b_i-1)$ required to reduce the number of bits $b_i$ on subcarrier $i$ by one bit is calculated according to (\ref{eq:BER}). The subcarrier, $i'$, with the maximum value of $\Delta \mathcal{P}_i$ is identified, its allocated bits $b_{i'}$ is set to $b_{i'}-1$, and its power is reduced by $\Delta \mathcal{P}_{i'}$. If the power constraint is not met, the process repeats. The proposed algorithm can be formally stated as follows. %shown in Table \ref{table:algo}.
%\vspace*{-0.25cm}
\floatname{algorithm}{}
\begin{algorithm}
\renewcommand{\thealgorithm}{}
\caption{\textbf{Proposed Algorithm}}
\begin{algorithmic}[1]
\State \textbf{INPUT} The AWGN variance $\sigma^2_n$, channel gain per subcarrier $i$ ($\mathcal{H}_i$), target BER per subcarrier $i$ ($\textup{BER}_{th,i}$), and weighting factor $\alpha$.
\For{$i$ = 1, ..., $N$}
\If{$\mathcal{C}_i \geq - \frac{4}{1.6} \: \frac{\alpha \ln(2)}{1-\alpha} \: \ln(5\:\textup{BER}_{th,i})$}
\State - $b_i^*$ and $\mathcal{P}_i^*$ are given by (\ref{eq:eq10}) and (\ref{eq:eq12}), respectively.
\State - $b_i^*$ $\leftarrow$ Round $b_i^*$ to the nearest integer.
\State - $\mathcal{P}_i^*$ $\leftarrow$ Recalculate $\mathcal{P}_i^*$ according to (\ref{eq:BER}).
\Else
\State - Null the corresponding subcarrier $i$.
\EndIf
\EndFor
\While{$\sum_{i = 1}^{N}\mathcal{P}_i > \mathcal{P}_{th}$}
\For{$i$ = 1, ..., $N$} \label{st:step_2}
\State - Calculate $\mathcal{P}_i(b_i-1)$ corresponding to reducing the number of bits $b_i$ on subcarrier $i$ to $b_i-1$, according to (\ref{eq:BER}). If $b_i-1 < 2$, null the subcarrier $i$.
\State - Calculate $\Delta \mathcal{P}_i = \mathcal{P}_i(b_i) - \mathcal{P}_i(b_i-1)$.
\EndFor
\State - Find subcarrier $i'$ with maximum $\Delta \mathcal{P}_i$.
\State - Set $b_{i'}$ to $b_{i'}-1$ and $\mathcal{P}_{i'}$ to $\mathcal{P}_{i'}(b_{i'}) - \Delta \mathcal{P}_{i'}$.
\EndWhile
\State \textbf{OUTPUT} The suboptimal $b_i$ and $\mathcal{P}_i$, $i$ = 1, ..., $N$.
\end{algorithmic}
\end{algorithm}

\section{Numerical Results} \label{sec:sim}

This section investigates the performance of the proposed algorithm, and compares it with that of the allocation scheme in \cite{wyglinski2005bit} and the exhaustive search for the discrete optimal allocation. The computational complexity of these algorithms is additionally compared.

\subsection{Simulation Setup}
We consider an orthogonal frequency division multiplexing (OFDM) system with a total of $N$ = 128 subcarriers. Without loss of generality, the BER constraint per subcarrier, $\textup{BER}_{th,i}$, is assumed to be the same for all subcarriers and set to $10^{-4}$. The channel impulse response $h(n)$ of length $N_{ch}$ is modeled as independent complex Gaussian random variables with zero mean and exponential power delay profile \cite{lin2008low}, i..e, $\mathbb{E}\{\left | h(n) \right |^2\} = \sigma_h^2 \: e^{-n\Xi}$, $n = 0, 1, ..., N_{ch}-1$,    
%\setlength{\arraycolsep}{0.0em}
%\begin{eqnarray}
%\mathbb{E}\{\left | h(n) \right |^2\} = \sigma_h^2 \: e^{-n\Xi}, \qquad n = 0, 1, ..., N_{ch}-1,
%\end{eqnarray}
where $\sigma_h^2$ is a constant chosen such that the average energy per subcarrier is normalized to unity, i.e., $\mathbb{E}\{\left | \mathcal{H}_i \right |^2\}$ = 1, and $\Xi$ represents the decay factor. Representative results are presented in this section and were obtained by repeating Monte Carlo trials for $10^{4}$ channel realizations with a channel length $N_{ch} = 5$ taps and decay factor  $\Xi = \frac{1}{5}$.

\subsection{Performance of the Proposed Algorithm}
\begin{figure*}[!t]
	\centering
		\includegraphics[width=1.00\textwidth]{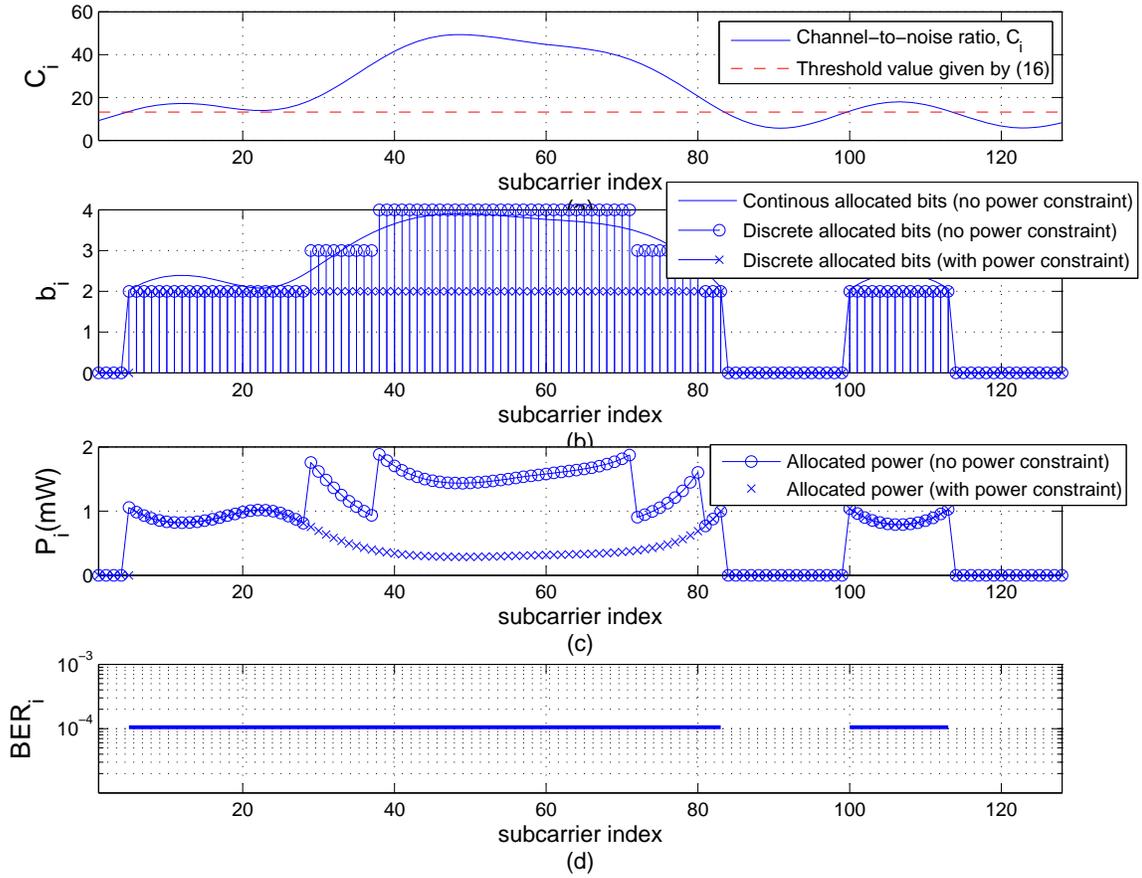}
	\caption{An example of the allocated bits and power per subcarrier for a given channel realization, with and without power constraint, at SNR = 10 dB, $\alpha$  = 0.5.}
	\label{fig:ch_realization}
\end{figure*}

Fig. \ref{fig:ch_realization} illustrates the allocated bits and powers with and without considering the total power constraint for an example channel realization, SNR = 10 dB and $\alpha$  = 0.5. Without considering the total power constraint, it can be seen from the plots in Fig.~\ref{fig:ch_realization} that when the channel-to-noise ratio per subcarrier, $\mathcal{C}_i$, exceeds the value in (\ref{eq:condition}), the number of bits and power allocated per subcarrier are non-zero. As expected, (\ref{eq:eq10}) yields a non-integer number of allocated bits per subcarrier, which is not suitable for practical implementations. This value is rounded to the nearest integer, as shown in Fig.~\ref{fig:ch_realization} (b), and the modified value of the allocated power per subcarrier to maintain the same $\textup{BER}_{th,i}$ is determined using (\ref{eq:BER}). When considering the total power constraint, for the sake of illustration, the power threshold is set to half the total transmit power with no power constraint; subcarriers with maximum $\Delta \mathcal{P}$ are identified, and the corresponding bits are reduced by one until the total power constraint is met, while guaranteeing the target BER requirement.    

\begin{figure}[!t]
	\centering
		\includegraphics[width=0.50\textwidth]{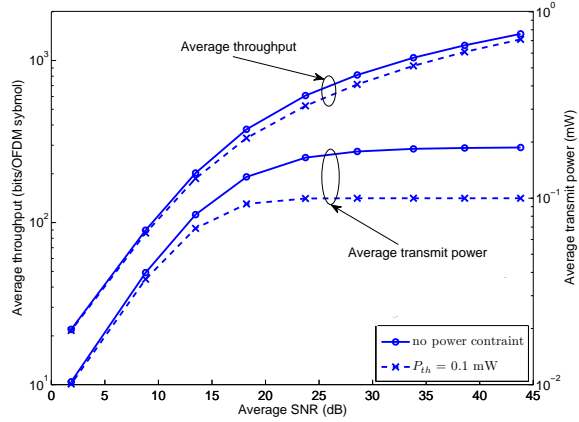}
	\caption{Average throughput and average transmit power as a function of average SNR, with and without power constraint, at $\alpha$  = 0.5.}
	\label{fig:T_P_vs_SNR}
\end{figure}

Fig. \ref{fig:T_P_vs_SNR} depicts the average throughput and average transmit power as a function of average SNR\footnote[4]{The average SNR is calculated by averaging the instantaneous SNR values per subcarrier over the total number of subcarriers and the total number of channel realizations, respectively.}, with and without considering the total power constraint at $\alpha$ = 0.5. Without considering the total power constraint and for an average SNR $\leq$ 24 dB, one finds that both the average throughput and the average transmit power increase as the SNR increases, whereas for an average SNR $\geq$ 24 dB, the transmit power saturates, and the throughput continues to increase. This observation can be explained as follows. For lower values of the average SNR, the corresponding values of $\mathcal{C}_i$ result in the nulling of many subcarriers in (\ref{eq:condition}). By increasing the average SNR, the number of used subcarriers increases, resulting in a noticeable increase in the throughput and power. Apparently, for SNR $\geq$ 24 dB, all subcarriers are used, and our algorithm essentially minimizes the average transmit power by keeping it constant, while increasing the average throughput. By considering a total power constraint, $\mathcal{P}_{th}$ = 0.1 $mW$, at lower SNR values when the total transmit power is below the threshold, the same average transmit power and throughput are obtained; however, at higher SNR values, when the total transmit power exceeds the threshold, a small reduction in the average throughput is noticed, which emphasizes that the proposed algorithm meets the power constraint while maximizing the throughput, i.e., the throughput does not degrade much when compared to the case of no power constraint.

\begin{figure}[!t]
	\centering
		\includegraphics[width=0.50\textwidth]{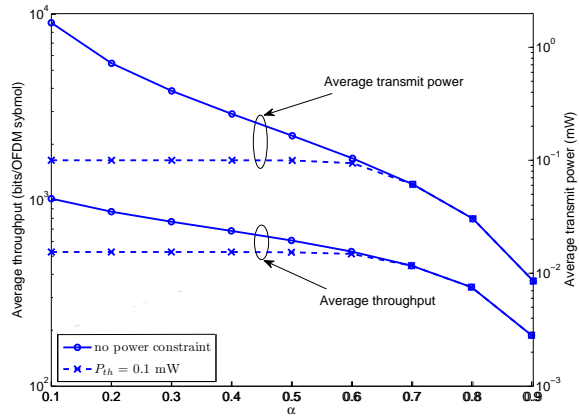}
	\caption{Average throughput and average transmit power as a function of $\alpha$, with and without power constraint, at $\sigma_n^2 = 10^{-3}$ $\mu W$.}
	\label{fig:proposed_alpha}
\end{figure}

Fig. \ref{fig:proposed_alpha} shows the average throughput and average transmit power as a function of the weighting factor $\alpha$, for $\sigma_n^2 = 10^{-3}$ $\mu W$, with and without considering the total power constraint. Without considering the total power constraint, one can notice that an increase of the weighting factor $\alpha$ yields a decrease of both the average throughput and average transmit power. This can be explained as follows. By increasing $\alpha$, more weight is given to the transmit power  minimization (the minimum transmit power is further reduced), whereas less weight is given to the throughput maximization (the maximum throughput is reduced), according to the problem formulation. By considering a total power constraint, $\mathcal{P}_{th}$ = 0.1 $mW$, the same average throughput and power are obtained if the total transmit power is less than $\mathcal{P}_{th}$, while the average throughput and power saturate if the total transmit power exceeds $\mathcal{P}_{th}$. Note that this is different from Fig.~\ref{fig:T_P_vs_SNR}, where the average throughput increases while the transmit power is kept constant, which is due the increase of the average SNR value. Fig. \ref{fig:proposed_alpha} illustrates the benefit of introducing such a weighting factor in our problem formulation to tune the average throughput and transmit power levels as needed by the wireless communication system.

\begin{figure}[!t]
	\centering
		\includegraphics[width=0.50\textwidth]{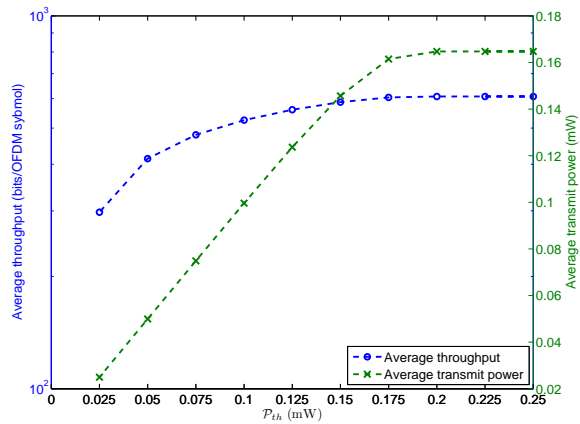}
	\caption{Average throughput and average transmit power as a function of the power constraint $\mathcal{P}_{th}$, at $\alpha$ = 0.5 and $\sigma_n^2 = 10^{-3}$ $\mu W$.}
	\label{fig:T_P_vs_power_constraint}
\end{figure}

In Fig. \ref{fig:T_P_vs_power_constraint}, the average throughput and average transmit power are plotted as a function of the power threshold $\mathcal{P}_{th}$, at $\alpha$ = 0.5 and $\sigma_n^2 = 10^{-3}$ $\mu W$. It can be noticed that the average throughput increases as $\mathcal{P}_{th}$ increases, and saturates for higher values of $\mathcal{P}_{th}$; moreover, the average transmit power increases linearly with $\mathcal{P}_{th}$, while it saturates for higher values of $\mathcal{P}_{th}$. This can be explained, as for lower values of $\mathcal{P}_{th}$, the total transmit power is restricted by this threshold value, while increasing this threshold value results in a corresponding increase in both the average throughput and  total transmit power. For higher values of $\mathcal{P}_{th}$, the total transmit power is always less than the threshold value, and, thus, it is as if the constraint on the total transmit power is actually relaxed. In this case, the proposed algorithm essentially minimizes the transmit power by keeping it constant; consequently, the average throughput remains constant for the same noise variance as for the previous scenario.

\subsection{Performance and Complexity Comparison}
\begin{figure}
	\centering
		\includegraphics[width=0.50\textwidth]{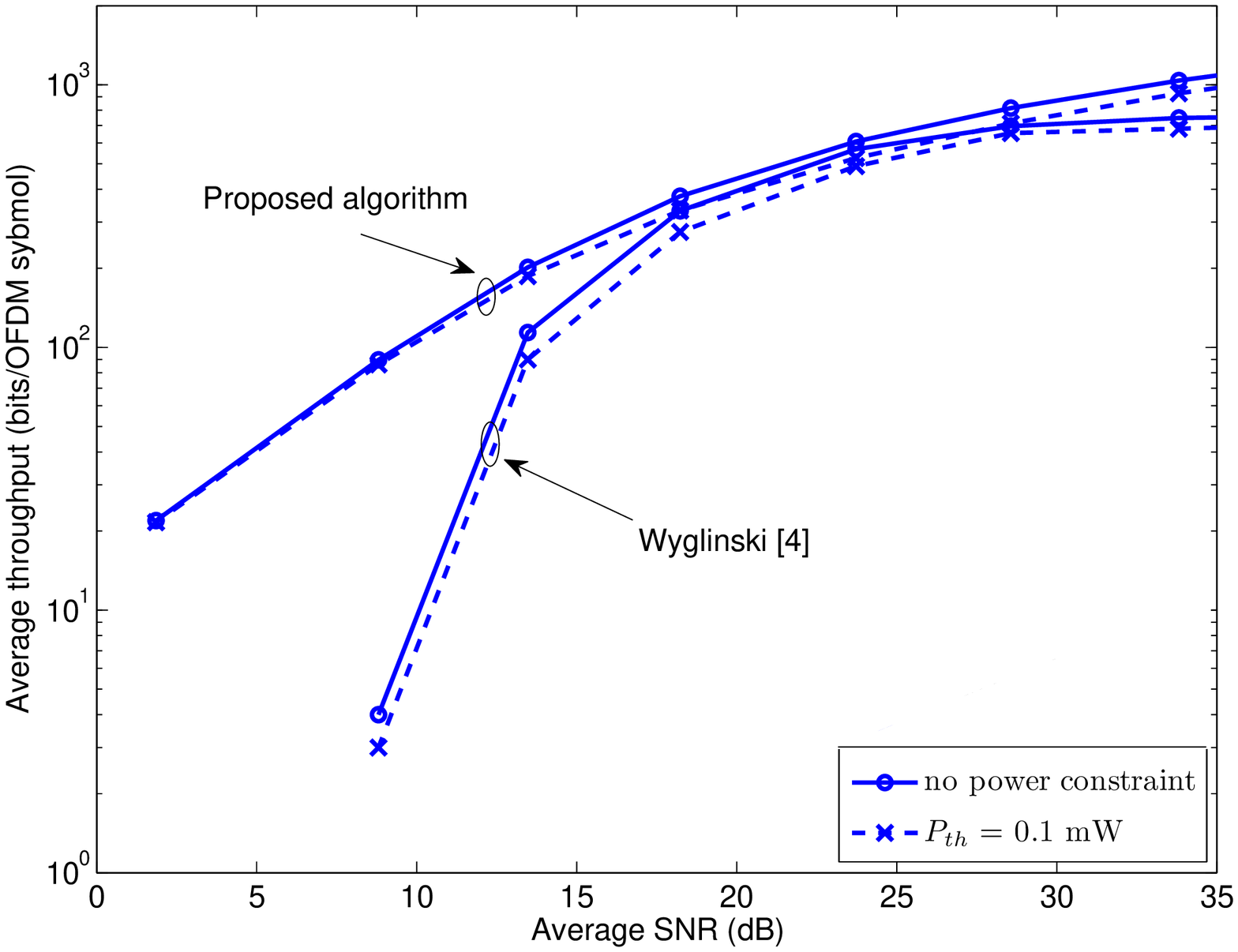}
	\caption{Average throughput as a function of average SNR for the proposed algorithm and Wyglinski's algorithm in \cite{wyglinski2005bit}, with and without power constraint, at $\alpha$ = 0.5.}
	\label{fig:Throughput_comp}
\end{figure}

In Fig. \ref{fig:Throughput_comp}, the throughput achieved by the proposed algorithm is compared to that obtained by Wyglinski's algorithm \cite{wyglinski2005bit} for the same operating conditions, with and without considering the total power constraint. For a fair comparison, the uniform power allocation used by the allocation scheme in \cite{wyglinski2005bit} is computed by dividing the average transmit power allocated by our algorithm by the total number of subcarriers. As shown in Fig.~\ref{fig:Throughput_comp}, the proposed algorithm provides a significantly higher throughput than the scheme in \cite{wyglinski2005bit} for low average SNR values. This result demonstrates that optimal allocation of transmit power is crucial for low power budgets.

%(at lower average SNR values, the power constraint is ineffective as discussed in Fig.~\ref{fig:T_P_vs_SNR}). This result emphasises the importance of adding the transmit power to the objective function while having a constraint on the total transmit power. 
%Furthermore, for increasing average SNR values, the average transmit power is constant as seen in Fig.~\ref{fig:T_P_vs_SNR} for values $\geq$ 24 dB, which in turn results in a saturating throughput for Wyglinski's algorithm. In contrast, the proposed algorithm provides an increasing throughput for the same range of SNR values. 
\begin{figure}[!t]
	\centering
		\includegraphics[width=0.50\textwidth]{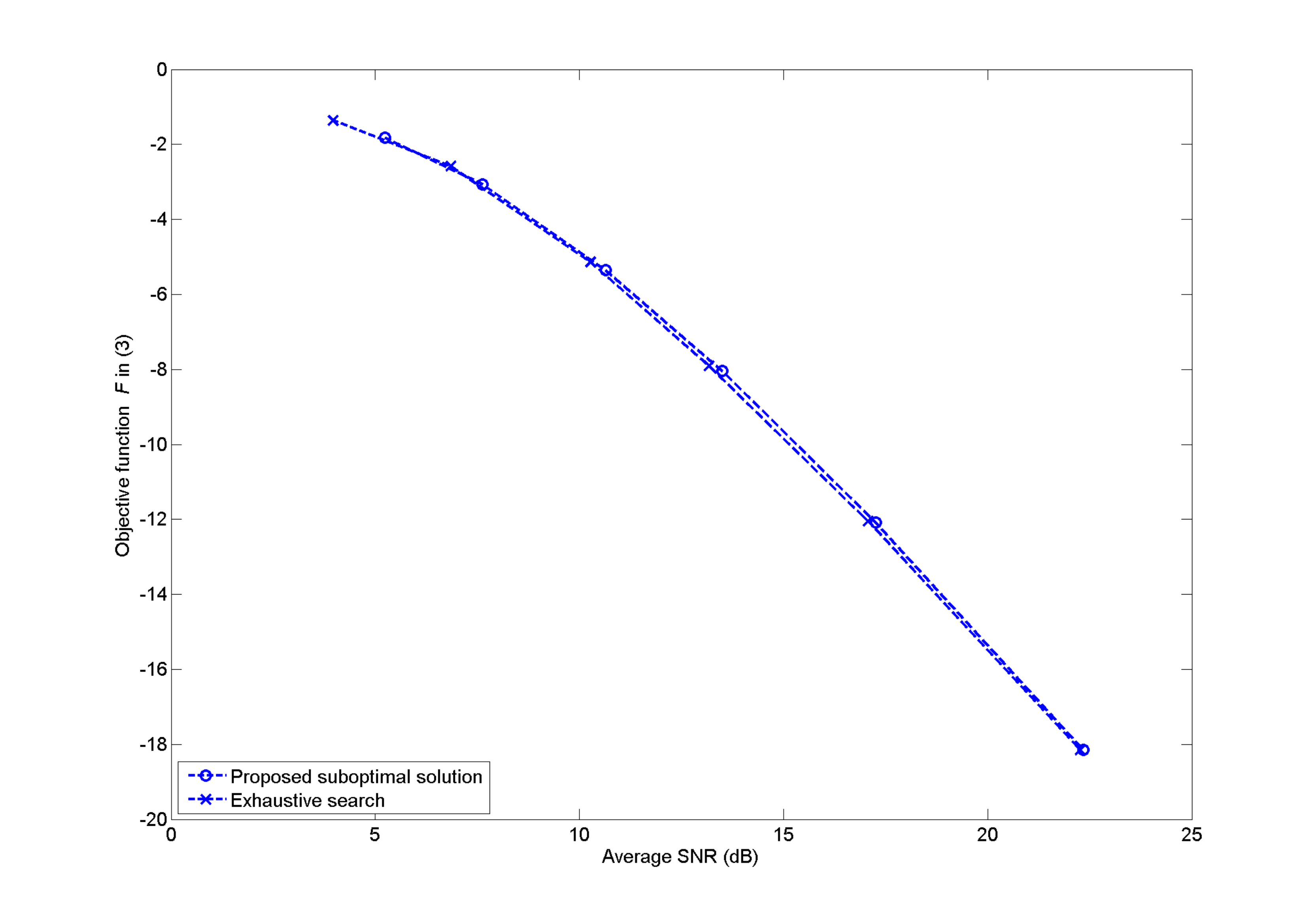}
	\caption{Objective function for the proposed suboptimal algorithm and the exhaustive search when $\mathcal{P}_{th}$ = 5 $\mu$W, $\alpha$ = 0.5 and $N$ = 8.}
	\label{fig:ex}
\end{figure}

%\vspace{-2pt}
To characterize the gap between the proposed suboptimal algorithm and the optimal solution, Fig. \ref{fig:ex} compares values of the objective function achieved with the proposed suboptimal algorithm and the optimal exhaustive search.  Note that the latter finds the discretized optimal allocation for the problem in (\ref{eq:ineq_const}). Results are presented for $\mathcal{P}_{th}$ = 5 $\mu$W, $\alpha$ = 0.5 and $N$ = 8; a small number of subcarriers is chosen, such that the exhaustive search is feasible. As can be seen in Fig. \ref{fig:ex}, the proposed suboptimal algorithm approaches the optimal results of the exhaustive search. 

Based on the algorithm description in Section \MakeUppercase{\romannumeral 2}-\textit{C}, one can show that the worst case computational complexity of the proposed algorithm is of $\mathcal{O}(\mathcal{N}^2)$ if the power constraint is effective, whereas it is of $\mathcal{O(N)}$ if the power constraint is ineffective, which is similar to or lower than the $\mathcal{O}(\mathcal{N}^2)$ of the Wyglinski's algorithm, and significantly lower than $\mathcal{O}(\mathcal{N}!)$ of the exhaustive search.

\section{Conclusion} \label{sec:conc}
\vspace{-5pt}
This paper proposed a novel suboptimal algorithm that jointly maximizes the total throughput and minimizes the total transmit power, with constraints on the BER per subcarrier and the total transmit power, for multicarrier communication systems. Simulation results demonstrated that the proposed algorithm outperforms the algorithm in \cite{wyglinski2005bit} under the same operating conditions, with similar or reduced computational effort. Additionally, it was shown that its performance approaches that of an exhaustive search with significantly lower complexity.
\vspace{-5pt}
%The proposed algorithm provided a reasonable close throughput when compared to the case of no constraint on the total transmit power.  Limiting the transmit power is a crucial requirement in wireless network environments, but this comes at the price of higher computational complexity due to the supplementary constraint on the power. In future work, we will seek other techniques,  with comparable complexity, including numerical solutions of this problem.

% or


%\appendices
%% you can choose not to have a title for an appendix
%% if you want by leaving the argument blank
%\section{}

\section*{Acknowledgment}

The authors would like to thank Dr. Howard Heys for his help on the analysis of the computational complexity of the recurrence relation in the proposed algorithm.

This work has been supported in part by the Communications Research Centre, Canada.%
\end{document}